\documentclass[reprint, amsmath, amssymb, aps, rmp, superscriptaddress]{revtex4-2}
\usepackage{format}
\allowdisplaybreaks

\begin{document}

\title{ Probability distributions of atomic scattering lengths  }

\author{ John L. Bohn }
\author{ Reuben R. W. Wang }
\affiliation{%
 JILA, NIST, and Department of Physics, University of Colorado, Boulder, Colorado 80309, USA 
}%

\date{\today}

 \begin{abstract}
 The probability distribution of the real and imaginary parts of atomic scattering lengths $a$ are derived, in a two-channel model that allows for inelastic scattering to occur.  While the real part of $a$ remains Cauchy-distributed, as predicted for single channel scattering in the classic work of Gribakin and Flambaum, the imaginary part of $a$ is seen to be strongly peaked near zero.  Two-body inelastic scattering  rates may therefore be smaller in general than a naive estimate would suggest.
 \end{abstract}

\maketitle

\section{Introduction}

This note serves as both a greeting to Ravi Rau and a dispatch from the world of dilute, ultracold gases, where  neutral atoms and molecules collide at typically sub-microKelvin temperatures.  Even though Ravi never explicitly published in this area, nevertheless his influence is strongly felt.  

The atoms and molecules in these gases collide at sufficiently low energy that they are in the Wigner threshold limit, hence their scattering is strongly dominated by the familiar threshold laws.  Ravi has been a tireless champion of threshold physics, summarized in his famous work with Fano \cite{Fano_Rau}, and in an influential review article \cite{Sadeghpour00_JPB}.  His classic treatment of the Wannier threshold law for double ionization highlights the ability of a single quantity, the exponent characterizing the energy dependence $E^{\alpha}$ of the process, to reveal information on detailed correlations of the charged particles \cite{Rau71_PRA}.

In the somewhat more pedestrian world of ultracold collisions of neutral atoms, the relevant Wigner threshold laws are well known.  Very typically, the collision is dominated by the lowest, $s$ partial wave, and the elastic scattering phase shift is linear in wave number, $\delta_0 = - ak$.  This $k$-dependence is standard; what varies from atom to atom, and what matters most in the context of ultracold gases, is the value of the prefactor, the scattering length $a$.  

At stake is the nature of the scattering cross section, which is responsible for bringing the gas to thermal equilibrium, and which therefore determines the ability to make an  ultracold gas at all.  Famously, the scattering length of $^{87}$Rb is approximately $a = 100 a_0$, $a_0$ being the Bohr radius.  This value is sufficiently large that evaporative cooling of this atom successfully led to the first Bose-Einstein condensate \cite{Anderson95_Sci}.  By contrast, the isotope $^{85}$Rb has a negative naturally-occurring scattering length \cite{Boesten97_PRA}.  This leads to an unfortunately-placed Ramsauer-Townsend minimum in its cross section, limiting evaporative cooling \cite{Burke98_PRL}.  (By various  devices, scattering lengths can be altered to necessary values, but that is a different story for another Fetschrift.)  

Scattering lengths tend to be extremely sensitive functions of the potential energy surface, meaning that even for alkali atoms, they cannot be predicted from {\it ab initio} theory (with the exception of one heroic recent result in \cite{Gronowski20_PRA}).  Generally, they are determined by thoughtful iterations of theory and experiment.  In this way, scattering lengths for most combinations of alkali atoms are now known, some to high precision.  To do so requires the evaluation of two scattering lengths, for the singlet and triplet Born-Oppenheimer potentials existing between these atoms.  By extension, three scattering lengths of higher-spin chromium atoms have also been extracted \cite{Werner05_PRL,Pavlovic05_PRA}, and have proven adequate for describing data.

Beyond this, the situation quickly becomes untenable.  Certain lanthanide atoms are perfectly amenable to laser cooling, yet their interactions are quite complex.  For example,  interactions of open-shell dysprosium atoms would require 81 distinct Born-Oppenheimer potentials, each with its own scattering length that presumably contributes to observed scattering \cite{Kotochigova11_PCCP}.  Extracting a quantitative model from data is, at present, considered inconceivable.  Even more challenging will be the equivalent procedure in collisions of ultracold molecules, which represents a rapidly growing area of endeavor.

Faced with the difficulties of detailed analysis, it may prove useful to consider instead trends that one could follow over the breadth of possibilities among many collision partners.  In the present note we will consider a statistical overview.  Statistics in ultracold collisions was introduced by Gribakin and Flambaum \cite{Gribakin93_PRA}, who derived, from semiclassical theory, the most likely value of the scattering length for long-range potentials that fall off as a power law, $-1/r^n$, of the distance $r$ between to atoms.  The true scattering lengths of various species should vary around this most-likely value, in such a way that, according to  Gribakin and Flambaum, three-quarters of all naturally occurring scattering lengths should be positive, for an ordinary van der Waals potential with $n=6$.  What these authors did not quite do (but surely could have) is to describe the full distribution function of scattering lengths.  In Sec. II we will complete this derivation, in preparation for the remainder of the article.

In addition to elastic scattering, it is extremely important to track inelastic scattering in the ultracold environment.  Even the smallest of atomic energy spacings, say hyperfine energies, are orders of magnitude larger than the translational temperature of the gas.  Thus an inelastic collision that releases this energy is a disaster: the products either leave the trap, or, perhaps worse, heat the remaining gas.  The atoms are like waiters in a busy restaurant, delicately balancing trays full of cocktails.  Should they collide, there will be a real mess.  

In the argot of cold collisions, these disruptive events are denoted by the technical term ``bad.''  Generally, it is accepted that any bad collisions that are allowed by energy conservation and symmetry considerations tend to happen at high collision rates, and should therefore be avoided if possible.  (Much ingenuity has gone toward finding ways to mitigate bad collisions, but again, this is not our story here.)  Exceptions of course exist.  In a serendipitous experiment at JILA, it was found that a mixture of $^{87}$Rb atoms in two distinct hyperfine states not only survived evaporative cooling, but could be simultaneously Bose-condensed \cite{Myatt97_PRL}.  The anomalously low inelastic spin-exchange rate that allowed this miracle was quickly understood to rely on an interference between singlet and triplet scattering, that is, on the near-coincidence of singlet and triplet scattering lengths for this isotope \cite{Kokkelmans97_PRA, Julienne97_PRL, Burke96_PRA}. 
Here was a statistical outlier.  

In the context of ultracold atoms, it is worthwhile to know how likely it is that such a calamitous event will occur.  That is, the question is one of probabilities.  To this end, in Sec. III we extend the Gribakin-Flambaum model to a two-channel case that allows for inelastic scattering.  We will cast the inelastic loss in terms of the imaginary part of the scattering length, and determine an approximate probability distribution for this quantity. 

To do so requires welding together the long-range physics that determines the threshold law, with the short-range physics that governs the change in state.  Here Ravi has also paved the way, stressing that multichannel quantum defect theory (MQDT) is an extremely versatile tool, far beyond its initial application to Rydberg atoms \cite{Fano_Rau}.  The ideas and notations that ground our theory in the following are rooted in the seminal work of Greene, Rau, and Fano \cite{Greene82_PRA}.

\section{Single Channel Scattering Lengths}

We first consider $s$-wave scattering in a single channel with potential $V(r)$, governed by the Schr\"odinger equation
\begin{align}
    \left( - \frac{ \hbar^2 }{ 2 m_r } \frac{ d^2 }{  dr^2 } + V \right)\psi 
    =
    E \psi,
\end{align}
where $m_r$ is the reduced mass of the collision partners.  For purposes of statistics, we envision an assembly of potentials $V$, collected from an ensemble of potential collision partners across the periodic table.  This variety can also include various Born-Oppenheimer curves for given partners, for example, the singlet and triplet curves of the alkalis, assumed to give scattering phase shifts independent of each other.  Different isotopes of the same element are not considered to have independent phase shifts as they are, to a good approximation, related by simple mass scaling \cite{Kitagawa08_PRA}. 

To include this variety of potentials as our ensemble, it is essential to reduce them to a common system of reduced units.  For threshold scattering, a relevant set of natural units is obtained from the long-range behavior.  In this note we restrict attention to those potentials with long-range van der Waals behavior characterized by the form $V(r) \approx -C_6/r^6$.  The corresponding natural unit of length is
\begin{align}
    r_6 
    =
    \left( \frac{ 2 m_r C_6 }{ \hbar^2 } \right)^{1/4}.
\end{align}
This scale tends to be of the order $ \approx 100 a_0$ for many atoms; for Rb it is $165 a_0$.  The short-range physics is not necessarily amenable to a simple scaling between species; indeed, this is where the joy of variety comes from.  Such a scaling will not be necessary in the QDT picture we employ.

In the spirit of quantum defect theory, we identify standard  solutions for the long-range potential, denoted ${\hat f}$ and ${\hat g}$.  These are given a useful standardized form in the {\it magnum opus} of Ruzic {\it et al.} \cite{Ruzic13_PRA}, which we follow throughout.  The functions are chosen so that the irregular function ${\hat g} \rightarrow 0$ as $r \rightarrow \infty$ in the zero-energy limit, a choice that maximizes the linear independence of ${\hat f}$ and ${\hat g}$ in numerical applications.  With this choice, the reference function ${\hat f}$ has phase shift $\eta = - {\bar a}k$ defined by the scattering length
\begin{align}
    {\bar a} =
    r_6
    \left( \frac{ \pi }{ 2^{3/2} \Gamma(5/4) \Gamma(1/2) } \right)^2 
    \approx 
    0.4780 \; r_6,
\end{align}
 which coincides exactly with the Gribakin-Flambaum most-likely scattering length \cite{Gribakin93_PRA}.  
 
The reference  functions ${\hat f}$ and ${\hat g}$ are related to the energy-normalized reference functions $f$ and $g$ in the usual way \cite{Greene82_PRA}:
\begin{align}
    \begin{pmatrix}
        f \\ g 
    \end{pmatrix}
    &= 
    \begin{pmatrix}
        A^{1/2} & 0 \\  A^{-1/2} {\cal G} & A^{-1/2} 
    \end{pmatrix}
    \begin{pmatrix}
        {\hat f} \\ {\hat g}
    \end{pmatrix}.
\end{align}
This defines two more QDT parameters, which Ruzic works out explicitly in the $s$-wave threshold limit:
\begin{subequations}
\begin{align}
    A^{1/2} 
    &=
    - ({\bar a}k)^{1/2}, \\
    {\cal G}  
    &=
    ({\bar a} k)^2 
    \left[ -1 + \frac{ 1 }{ 3 } \left( \frac{ r_6 }{ {\bar a} } \right)^2 \right].
\end{align}
\end{subequations}

The statistical model is derived in QDT as follows.  For a given potential $V$, one would solve the Schr\"odinger equation, matching its solution $\psi$ to the reference functions at a convenient radius $r=r_0$,
\begin{align}
    \psi = {\hat f} - {\tilde K} {\hat g}.
\end{align}
This would define the short-range $K$-matrix
\begin{align} \label{eq:quantum_defect}
    {\tilde K} = \tan( \pi \mu )
\end{align}
in terms of a quantum defect $\mu$.   In the statistical model we do not consider any explicit potential $V$, but rather {\it assume} that the quantum defects from such a process would be uniformly distributed on the interval $\mu \in [-1/2, 1/2]$.  In this way the vast differences in depth and shape of the potentials for many different atoms are rendered irrelevant.  Whatever the atoms are actually doing down there, the net result is always encapsulated in the quantum defect  $\mu$.

By the rules of QDT, one then constructs the short-range phase shift $\delta_{sr}$ via
\begin{align}
    \tan \delta_{sr}  
    =
    \frac{ A^{1/2} {\tilde K} A^{1/2} }{ 1 + {\cal G}{\tilde K} }  \approx {\tilde K} {\bar a}k,
\end{align}
here ignoring ${\cal G}{\tilde K}$ as small compared to unity.  This quantity will become relevant if and when we consider the effective range.  
The physical scattering phase shift is given by the sum of long- and short-range contributions, 
\begin{align}
    \delta_0 = \eta + \delta_{sr} 
    \approx
    (-1 + {\tilde K}) {\bar a}k,
\end{align}
whereby the scattering length in units of ${\bar a}$ is given by
\begin{align}
    \frac{ a }{ {\bar a} } 
    =
    - \frac{ 1 }{ {\bar a} k } \delta_0 = 1-{\tilde K}.
\end{align}
 
To find the distribution of scattering lengths, we begin with the distribution of quantum defects,
\begin{align}
    P(\mu) 
    =
    \begin{cases}
        1, & -\frac{ 1 }{ 2 } \le \mu \le \frac{ 1 }{ 2 } \\
        0, & \rm{otherwise} 
    \end{cases}.
\end{align}
This assumption along with Eq.~(\ref{eq:quantum_defect}) implies a distribution of short-range $K$-matrices related to the former by 
\begin{align}
    P({\tilde K})   
    =
    \frac{ 1 }{ \pi } \frac{ 1 }{ {\tilde K}^2 + 1 }.
\end{align}
Thus the short-range $K$-matrix is distributed as a Lorentzian or, in the language of probability theory, a Cauchy distribution.  (That the tangent of a uniform distribution yields a Cauchy distribution is well-known.  Nevertheless, this result and all the others that we use below are derived in the Appendix.)  

Restoring the units, the distribution of scattering lengths is given by
\begin{align} \label{eq:a_dist}
    P(a) 
    =
    \frac{ 1 }{ \pi  } \frac{ {\bar a} }{ (a - {\bar a})^2 + {\bar a}^2 }.
\end{align}
Significantly, the Cauchy distribution has neither a well-defined mean nor a well-defined standard deviation.  It is, rather, characterized by its mode (most likely value) and its full-width at half-maximum (FWHM), both of which are ${\bar a}$ in this case.   From this distribution we evaluate  the fraction of scattering lengths that are positive,
\begin{align}
    \int_0^{\infty} da P(a) 
    =
    \frac{ 3 }{ 4 },
\end{align}
just as prophesied by Gribakin and Flambaum.  Having this distribution, we can say other things, for example, half of all scattering lengths should lie within ${\bar a}/2$ of the mode ${\bar a}$.

\section{Two Channels: Scattering and Loss}

Inelastic collisions that lead to bad outcomes are somewhat less universal than potential scattering, and in principle depend on the mechanism by which channel coupling occurs.  Nevertheless, for the kinds of collisions envisioned here, this mechanism may be assumed to lie at short range and to be subsumed in the short-range $K$-matrix, regarded as a set of parameters of the theory.  We therefore disregard, e.g., collisions of dipolar molecules, where torques exerted by the dipoles when they are far apart can drive inelastic scattering at large $r$ \cite{Avdeenkov02_PRA}

\subsection{Model and QDT}

For the sake of simplicity, we consider a two-channel system, where the incident channel 1 is at threshold, while the other channel 2 is exothermic by some energy much greater than the collision energy; in particular it is not at threshold.  The long-range potentials in both channels are assumed to scale as $-C_6/r^6$, whereby the QDT functions are computed for each channel as above.  In the incident channel near threshold, $\eta_1 = -{\bar a}k$, $A_1^{1/2} = -({\bar a}k)^{1/2}$, and as above we will not concern ourselves with ${\cal G}_1$.  In the outgoing channel which is far from threshold, $A_2^{1/2}=1$ and the values of $\eta_2$ and ${\cal G}_2$ are irrelevant. 

These two asymptotic channels are presumed to become coupled at short range, in a way that is well-approximated by a frame transformation.    It is assumed that the short-range physics is described by two alternative channels, each with its own quantum defect $\mu_{\lambda}$.  In this approximation the short-range $K$-matrix is diagonal in the short-range basis and has the form
\begin{align}
    {\tilde K}^{sr} 
    =
    \begin{pmatrix}
        \tan( \pi \mu_1 ) & 0 \\ 0 & \tan( \pi \mu_2 )
    \end{pmatrix}. 
\end{align}
Significantly, we do not perform the usual MQDT step of eliminating closed channels, as there are none in this example.  It should be remembered that we seek here the statistics of scattering lengths away from resonances.

In this $2 \times 2$ example the transformation between basis sets is a simple rotation through an angle $\theta$.  For any given collision, the value of $\theta$ will be determined by exactly what the short-range and asymptotic channels are, including the spin structure of the atoms and the mixing of channels by ambient electromagnetic fields.  To simplify the treatment we do not consider these details and assume that, across the ensemble of species and conditions considered, $\theta$ is uniformly distributed in $\theta \in [-\pi / 2, \pi /2]$.  

Expressed in the asymptotic basis, the short-range $K$-matrix in this notation then becomes
\begin{widetext}
\begin{align}
    {\tilde K} 
    =
    \begin{pmatrix}
        \cos^2 \theta \tan( \pi \mu_1) + \sin^2 \theta \tan( \pi \mu_2) 
        & \cos \theta \sin \theta [ \tan(\pi \mu_2) - \tan(\pi \mu_1) ] \\
        \cos \theta \sin \theta [ \tan(\pi \mu_2) - \tan(\pi \mu_1) ] 
        & \sin^2 \theta \tan(\pi \mu_1) + \cos^2 \theta \tan(\pi \mu_2) 
     \end{pmatrix}.
 \end{align}
\end{widetext}
This leads to the asymptotic $K$-matrix 
 \begin{align}
     K 
     &= 
     A^{1/2} {\tilde K} A^{1/2} \nonumber\\ 
     &=
     \begin{pmatrix}
         {\bar a}k {\tilde K}_{11} & - ({\bar a}k)^{1/2} {\tilde K}_{12} \\
         - ({\bar a}k)^{1/2} {\tilde K}_{21} & {\tilde K}_{22}
     \end{pmatrix},
 \end{align}
 followed by the $S$-matrix,
 \begin{align}
     S  
     =
     \begin{pmatrix}
         e^{-i{\bar a}k} & 0 \\ 0 & e^{i \eta_2}
     \end{pmatrix}
     \left( I + iK \right) \left( I - iK \right)^{-1}
     \begin{pmatrix}
         e^{-i{\bar a}k} & 0 \\ 0 & e^{i \eta_2}
     \end{pmatrix}.
 \end{align}
 
Writing the resulting phase shift in channel 1 as $S_{11} = \exp( 2i \delta_1)$, we define the complex scattering length in this channel via
\begin{align}
    a 
    =
    {\bar a}(\alpha - i \beta )
    =
    - \frac{ 1 }{ k } \delta_1.
\end{align}
This defines the dimensionless quantities $\alpha$ and $\beta$, regarded as real and imaginary parts of the scattering length in units of ${\bar a}$.
Expanding $S_{11}$ to linear order in $k$, these quantities are given by
\begin{subequations}
\begin{align}
    \alpha 
    &=
    1 - {\tilde K}_{11} 
    + \frac{{\tilde K}_{12}^2  }{  {\tilde K}_{22}^2 +1  } {\tilde K}_{22}, \label{subeq:real_scattering_length} 
    \\
     \beta  
    &= 
    \frac{ {\tilde K}_{12}^2 }{ {\tilde K}_{22}^2 + 1 }. \label{subeq:imag_scattering_length}
\end{align}
\end{subequations}

\subsection{Probability Distributions}

The scattering observables $\alpha$ and $\beta$ are  functions of the fundamental parameters of the model, $\mu_1$, $\mu_2$, $\theta$, which are treated as random variables.  By the standard formalism for transforming and composing random variables, one can then find the probability distributions for $\alpha$ and $\beta$.  These transformations are carried out in detail in the Appendix.  Here we present and explore the results.  For purposes of illustration, we have run a simulation choosing $10,000$ triples $(\mu_1,\mu_2,\theta)$ from their uniform distributions.  The subsequent quantities of the theory can then be calculated and displayed as histograms.

\begin{figure}[ht]
    \centering
    \includegraphics[width=\columnwidth]{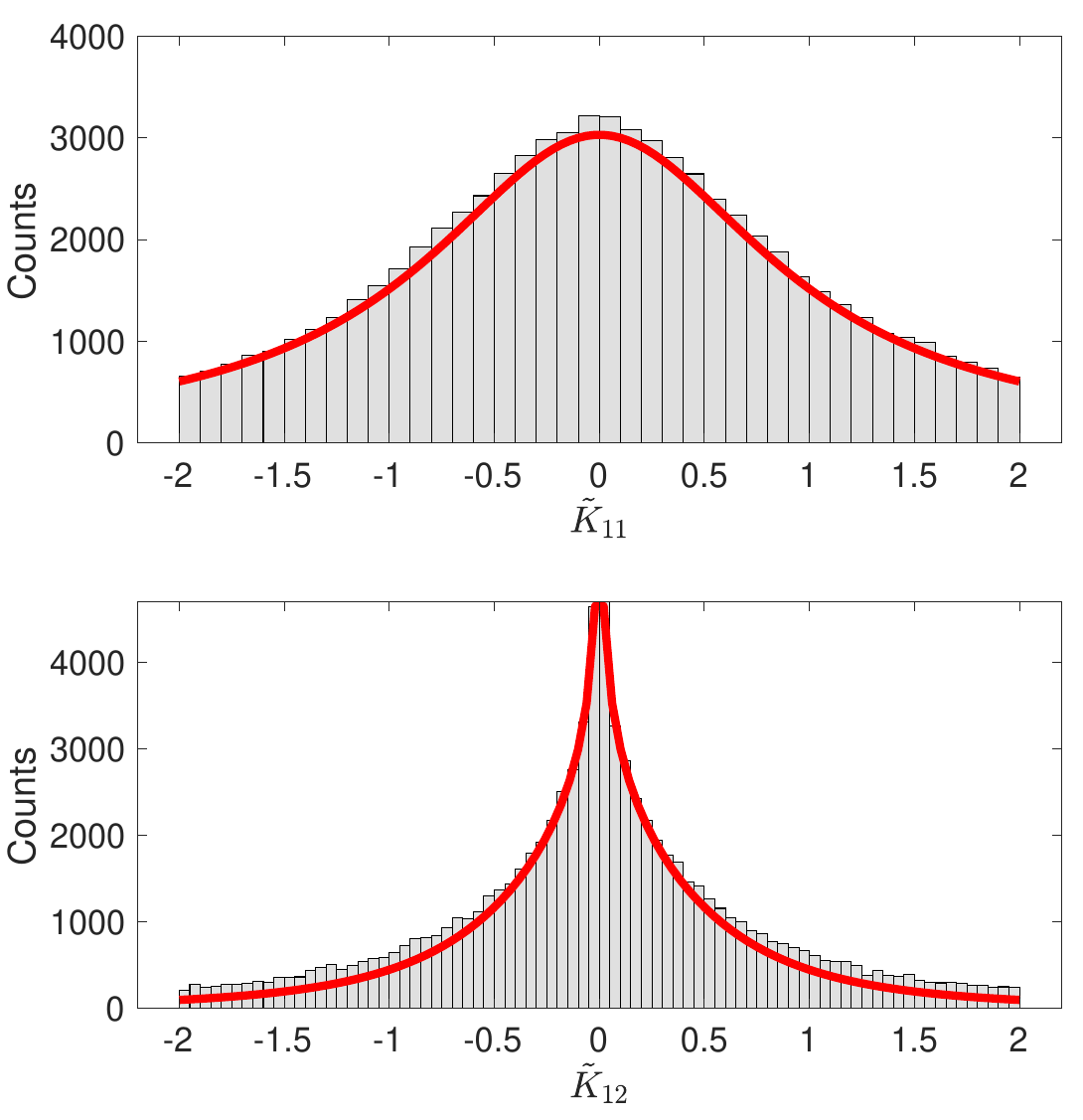}
    \caption{Probability distributions of the diagonal (upper) and off-diagonal (lower) elements of the short-range $K$-matrix.  In each case, the histogram is numerically sampled from the model in the text.  The red curves are the analytical formulas for the distributions, given in (\ref{eq:K_dist}).  The analytical curves are re-normalized to give the same integral as the histogram over the range shown.}
    \label{fig:Fig1}
\end{figure}

We begin with the distribution of elements of the short-range $K$-matrix, ${\tilde K}$.  The model predicts that these are distributed according to
\begin{subequations} \label{eq:K_dist}
\begin{align}
      P({\tilde K}_{11}) 
      &= 
      \frac{ 1 }{ \pi } \frac{ 1 }{ {\tilde K}_{11}^2 + 1 }, \label{subeq:K11_dist} 
      \\
      P({\tilde K}_{12}) &=\frac{ 2 }{ \pi^2 } \frac{ 1 }{ \sqrt{ ({\tilde K}_{12})^2 + 1} }
     \sinh^{-1}\left( \frac{ 1 }{ |{\tilde K}_{12}| } \right). \label{subeq:K12_dist}
\end{align}
\end{subequations}
Histograms of the numerical simulations of these quantities are plotted in Fig.~\ref{fig:Fig1}, along with (red lines) the formulas in (\ref{eq:K_dist}).  In the upper panel we see that the distribution of the diagonal matrix element ${\tilde K}_{11}$ is very well-described by the ordinary Cauchy distribution from the one-channel case.  While not shown, the distribution of ${\tilde K}_{22}$ is the same.    Each diagonal matrix element is the weighted sum of variables $\tan( \pi \mu_{\lambda})$ that are Cauchy-distributed.  The weights add to unity, whereby the average is also Cauchy distributed.  This is shown in detail in the Appendix.

More interesting, and somewhat unexpected, is the distribution of off-diagonal elements shown in the lower panel.  This distribution is far more strongly peaked near zero than the Cauchy distribution, a result captured in the analytical formula (\ref{subeq:K12_dist}).  The most likely value of ${\tilde K}_{12}$ is zero, but a FWHM is not possible to define here, as the distribution suffers a logarithmic divergence:
\begin{align} 
    \lim_{{\tilde K}_{12} \to 0} P({\tilde K}_{12}) 
    =
    \frac{ 2 }{ \pi^2 } \ln \left( \frac{ 2 }{ {\tilde K}_{12} } \right).
\end{align}
One can, however, make the following comparison.  For the Cauchy distribution that defines $P({\tilde K}_{11})$, half the distribution lies within $\pm {\bar a}$ of zero; for the distribution $P({\tilde K}_{12})$, half the distribution is within $\pm 0.55 {\bar a}$.  Thus in spite of the divergence, ${\bar a}$ is  still a relevant scale on which to consider the distribution.

\begin{figure}[ht]
    \centering
    \includegraphics[width=\columnwidth]{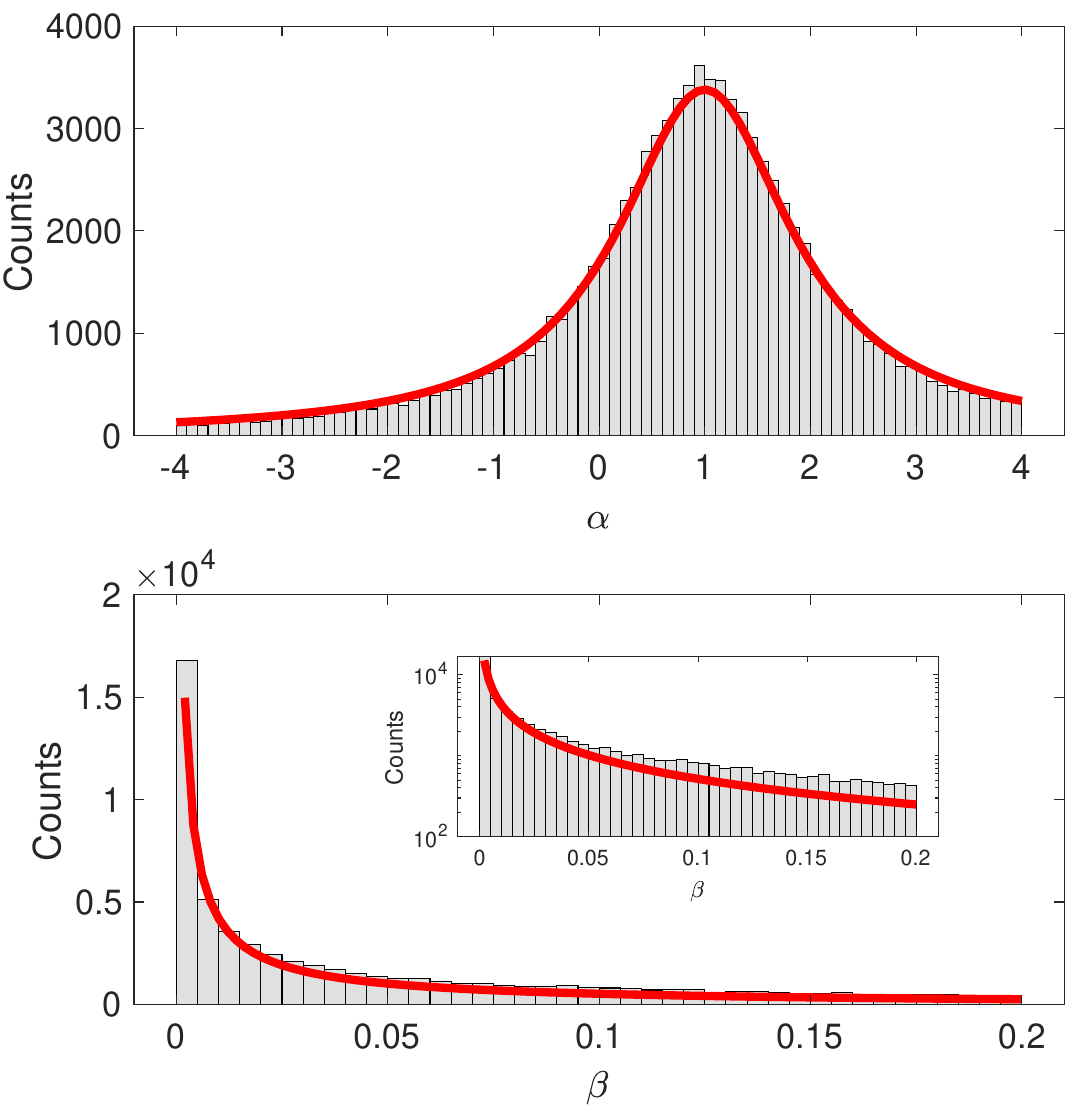}
    \caption{Probability distributions of the real (upper) and imaginary (lower) parts of the normalized scattering length  $a/{\bar a} = \alpha - i \beta$.    In each case, the histogram is numerically sampled from the model in the text.  The red curves are the analytical formulas for the distributions, given in (\ref{eq:alpha_beta_dist}).  The inset in the lower panel represents the same data, but with the vertical axis on a logarithmic scale, to better show the tail of the distribution.}
    \label{fig:Fig2}
\end{figure}

We now turn to the final results, the distributions of dimensionless real and imaginary parts of the scattering length.  These are displayed in Fig.~\ref{fig:Fig2}, with $\alpha$ in the upper panel and $\beta$ in the lower, and are compared to the approximate analytical formulas
\begin{subequations}
\label{eq:alpha_beta_dist}
\begin{align}
    P(\alpha) 
    &= 
    \frac{ 1 }{ \pi } \frac{ 1 }{ (\alpha - 1)^2 + 1 } \label{subeq:alpha_dist}
    \\
    P( \beta ) 
    &=
    \frac{ 1 }{ \pi^2 } \frac{ 1 }{ \sqrt{ \beta } } 
    \left[ \sinh^{-1} \left( \frac{ 1 }{ \sqrt{ \beta} } \right) \right]^2, \label{subeq:beta_dist}
\end{align}
\end{subequations}
These two formulas are the main result of this note.
  
The real part, $\alpha$ is well-described by the same Cauchy distribution (\ref{eq:a_dist}) as for the single channel case.  The reason for this is clear from the formula (\ref{subeq:K11_dist}).  The main contribution to $\alpha$ is given simply by $1 - {\tilde K}_{11}$, whereby the result follows trivially just as in the one-channel case.  The correction to this result, the second term of (\ref{subeq:real_scattering_length}), is proportional to ${\tilde K}_{12}^2$, hence is heavily peaked around zero and changes the scattering length but little.  In practice, this works out so that (\ref{subeq:alpha_dist}) is an excellent approximation. We conclude that, away from resonances, the two-channel elastic scattering length is distributed the same as a single-channel scattering length.

As for the imaginary part $\beta$, it is by its nature strictly non-negative, and is distributed sharply near zero, an expected behavior it inherits from ${\tilde K}_{12}$.  The analytical formula for the distribution is approximate, but seems to describe the peak at zero quite well.  The inset in the lower panel of Fig.~2 is the same histogram, but plotted with counts on a logarithmic scale, to better emphasize the tail of the distribution.  As can be seen, the formula somewhat underestimates the true distribution at large values of $\beta$, but we will not concern ourselves with this detail.

\section{Discussion}

Within  the model presented, a message stands out.  In the case of scattering in a single potential, we know that the real part of the scattering length is Cauchy distributed as given above, and that the imaginary part is rigorously zero.  The present results note that, if an additional channel is added into which scattering can occur, the real part of the scattering length remains Cauchy distributed, while the imaginary part still {\it tries very hard} to remain close to zero.  

It is not hard to imagine that the result for elastic scattering generalizes.  Consider scattering in some asymptotic channel $i$ in a multichannel system.  Within the frame transformation approximation assumed in this model, the non-resonant $K$-matrix in this channel will be given by the weighted average of diagonal $K$-matrices in each of the short-range channels $\lambda$:
\begin{align}
{\tilde K}_{ii} = \sum_{\lambda} \langle i | \lambda \rangle \tan(\pi \mu_{\lambda}) \langle \lambda | i \rangle.
\end{align}
And since the sum of squares of the coefficients of transformation is unity, we again recollect the Cauchy distribution from the single-channel case.  Thus the Gribakin-Flambaum result is generalized to non-resonant multichannel scattering.

Finally, let us put the result for the imaginary part of the scattering length into practical terms.  The role of $\beta$ is to track the flux that enters in channel 1 but departs in channel 2.  Using the unitarity of the $S$-matrix,
\begin{align}
    |S_{12}|^2 
    =
    1 - |S_{11}|^2 
    =
    1 - |\exp(-2 {\bar a}k \beta)|^2 \approx 4 {\bar a} k \beta,
\end{align}
giving the collision rate constant for inelastic collisions (regarded as bad),
\begin{align}
    {\cal K}_{\rm bad} 
    =
    g v \frac{ \pi }{ k^2 } |S_{12}|^2 
    = 
    g \left( \frac{ \hbar k }{ m_r } \right) \frac{ \pi }{ k^2 } (4 {\bar a} k \beta )
    = 
    g \frac{4  \pi \hbar }{ m_r } {\bar a} \beta.
\label{eq:K_bad}
\end{align}
Here $v = \hbar k / m_r$ is the collision velocity, and $g$ is a factor that accounts for symmetrization:  $g=1$ unless the  initial channel contains two identical atoms in identical internal states, in which case $g=2$.  In the event that these bad collisions lead to loss from the trap, their number density $n$ diminishes in time according to
\begin{align}
    \frac{ dn }{ dt } 
    =
    - {\cal K}_{\rm bad} n^2,
\end{align}
assuming that the loss is dominated by two-body scattering events.  

\begin{figure}[ht]
    \centering
    \includegraphics[width=\columnwidth]{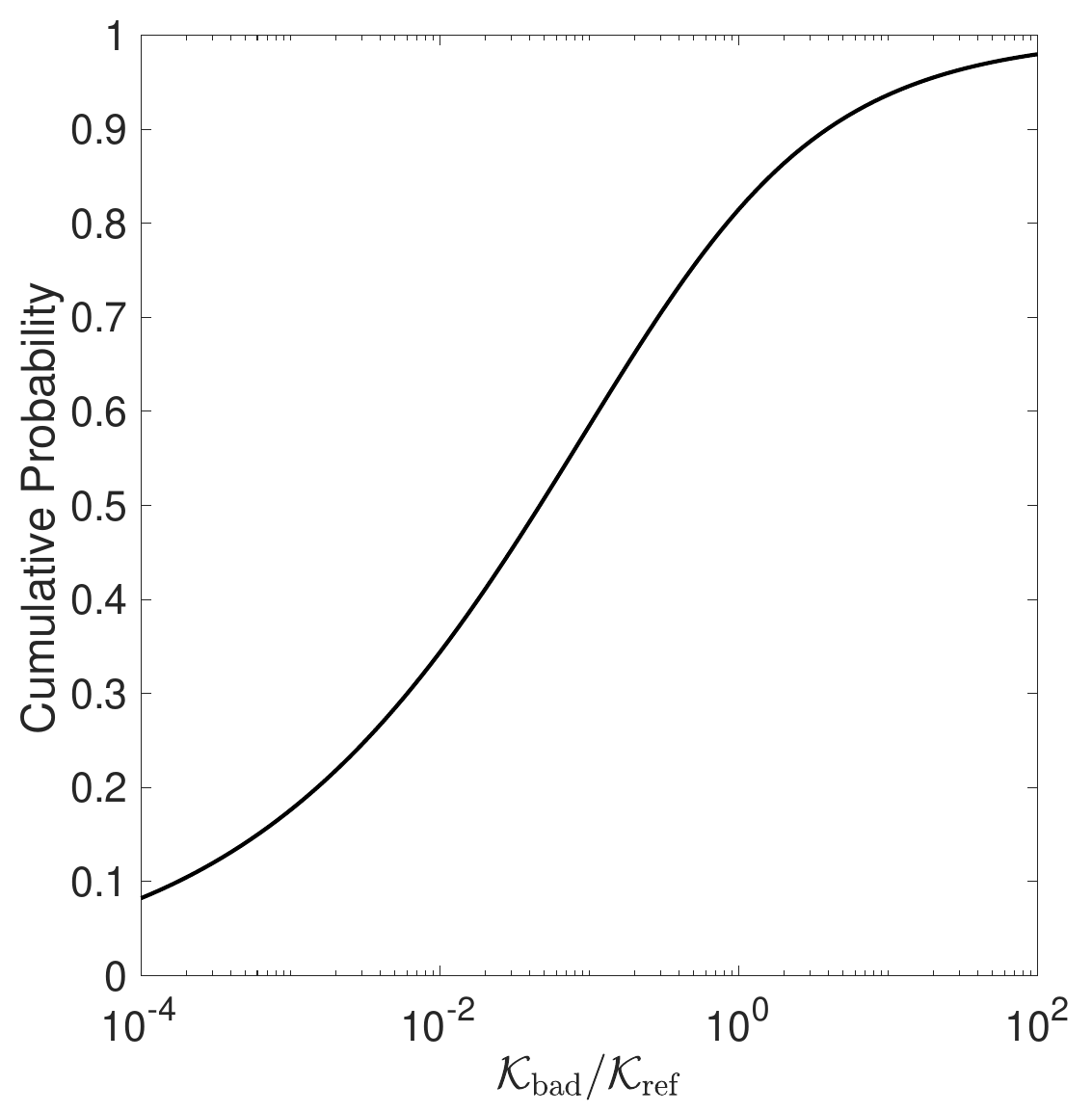}
    \caption{Cumulative probability distribution for the rate constant ${\cal K}_{\rm bad}$ for ``bad'' collisions, normalized by the reference rate ${\cal K}_{\rm ref} = g(4  \pi \hbar / m_r ) {\bar a}$.}
    \label{fig:Fig3}
\end{figure}

To put the result into perspective, consider the following.  Suppose you are building a new laboratory to cool and trap an atomic or molecular species that has not been trapped before, so that nothing is known about its collision properties.  (I do not think this is something Ravi is likely to do, but one never knows!)  Suppose, further, that some kind of bad collision process is possible, and that it occurs at short range.  This may include spin-exchange for atoms, chemical reactions for molecules, or perhaps light-assisted collisions for either.  In the context of collisions, all you know are the reduced mass and the $C_6$ coefficient, which can often be estimated in perturbation theory.

From this, you would like some sense of the size of the rate constant for bad collisions. You can construct a typical scale for this quantity by disregarding the influence of $\beta$, thus defining a reference rate constant
\begin{align}
    {\cal K}_{\rm ref} 
    =
    g \frac{4  \pi \hbar }{ m_r } {\bar a}.
\end{align}
In terms of this reference value, the true rate constant will be given by
\begin{align}
    {\cal K}_{\rm bad} 
    =
    {\cal K}_{\rm ref} \; \beta .
\end{align}
That is, the values of ${\cal K}_{\rm bad}$, in units of the reference value ${\cal K}_{\rm ref}$, are distributed just as the value of $\beta$ is in Eqn.~(\ref{subeq:beta_dist}).  

In this spirit, we present in Fig.~\ref{fig:Fig3} the cumulative probability distribution for the normalized bad rate constant.  From this figure we read that there is an approximately 80\% probability that the actual rate constant is smaller than ${\cal K}_{\rm ref}$; the easiest estimate is likely an over-estimate.  Even better: the odds are about 34\% that the actual rate constant is 100 times smaller than ${\cal K}_{\rm ref}$, thus bad scattering has at least a fighting chance of not being as bad as feared.  This is the ultimate consequence of the peaking of $\beta$ around zero.

To return to the context of the mixed-BEC experiment in \cite{Myatt97_PRL}, for rubidium we expect a reference rate constant of ${\cal K}_{\rm ref, Rb} = 7.7 \times 10^{-11}$ cm$^3$/s. The observed value, ${\cal K} = 2.2 \times 10^{-14}$ cm$^3$/s, is 3500 times smaller.  From our simple theory, finding a rate this small or smaller is an event with probability $\approx 12$\%.

The simple distributions presented here are of course subject to assumptions of the model. For example, they refer to scattering with only a single loss channel.  More significantly, the result assumes that the rotation angle $\theta$ is uniformly distributed. Nonetheless, the results are emblematic of future possibilities, where statistical understanding of ultracold collisions can be explored  through the lens of MQDT. 

This work is supported by the National Science Foundation under Grant No. PHY2110327.  JLB gratefully acknowledges advice and encouragement from Ravi Rau over the years, particularly in graduate school when things may have gone off the rails.

 \appendix
 
 \section{Transformation of Probability Distributions}
 
 Given certain variables with defined probability distributions functions (pdfs), it is a standard matter to find the pdfs of combinations of these variables.  The results used here are as follows.  Suppose $X$ is a random variable with probability distribution $P_X(x)$.  We now change variables to a new random variable $Y=Y(X)$, given as a function of the original.  Then the new pdf is
 \begin{align}
 P_Y(y) = P_X(x) \Big| \frac{ dy }{ dx } \Big|^{-1},
 \end{align}
 where on the right the inversion $x=x(y)$ is implied.

Given two pdfs, $P_X(x)$, $P_Y(y)$, assumed to by independent, the pdf of their sum $Z=X+Y$ is given by
\begin{align}
P_Z(z) &= \int dx \int dy P_X(x) P_Y(y) \delta( x + y - z) \\
&= \int dx P_X(x) P_Y(z-x),
\end{align}
while the pdf of their product $W=XY$ is given by
\begin{align}
P_W(w) & = \int dx \int dy P_X(x) P_Y(y) \delta( w - xy) \\
&= \int dx P_X(x) P_Y\left( \frac{ w }{ x } \right) \frac{ 1 }{ |x| }.
\end{align}
In both cases the limits of integration are those appropriate to the ranges of the original pdfs.  In practice, we evaluate these integrals in Mathematica, at least up  to the point where the resulting expression, even if in principle analytic, is no longer useful to look at.  
In the following we will omit the subscript on $P$, the random variable being assumed identified by the argument. 

For example, we have a quantum defect distributed according to $P(\mu) = 1$, for $\mu \in [-1/2,1/2]$, the corresponding $K$-matrix ${\tilde K} = \tan( \pi \mu)$ has pdf
\begin{align}
    P({\tilde K}) 
    &=
    P\Big(\mu({\tilde K})\Big) \left| \frac{ d{\tilde K} }{ d \mu } \right|^{-1} \nonumber\\
    &= 
    \frac{ 1 }{ \pi } \left( \frac{ 1 }{ \sqrt{1+{\tilde K}^2 } } \right)^2 \nonumber\\
    &=
    \frac{ 1 }{ \pi }  \frac{ 1 }{ {\tilde K}^2 +1 }.
\end{align}
  
Next we construct the pdf for the short-range $K$-matrix.  For example,
\begin{align}
{\tilde K}_{11} = \cos^2 \theta \tan(\pi \mu_1) + \sin^2 \theta \tan(\pi \mu_2).
\end{align}
Each random variable $t_i = \tan \pi \mu_i$  is Cauchy distributed.   Scaling these variables to, for example, $t_a=at$ yields the distribution 
\begin{align}
    P(t_a) 
    =
    \frac{ 1 }{ \pi }  \frac{ |a| }{ t_a^2 + a^2 }, 
\end{align}
with FWHM  $|a|$.  Thus, if $t_a = a \tan(\pi \mu_1)$ and $t_b = b \tan(\pi \mu_2)$ are two such scaled variables, their sum $t_{ab} = t_a+t_b$ has distribution
\begin{align}
    P(t_{ab}) 
    &= 
    \int_{-\infty}^{\infty} dt_a \frac{ 1 }{ \pi } \frac{ |a| }{ t_a^2 + a^2 }
    \frac{ 1 }{ \pi } \frac{ |b| }{ (t_{ab}-t_a)^2 + b^2 } \nonumber\\
    &= 
    \frac{ 1 }{ \pi } \frac{ |a| + |b| }{ z^2 + (|a|+|b|)^2 }.
    \label{eq:combined_Cauchy}
\end{align}
From this it follows that, for our matrix element ${\tilde K}_{11}$, with $a = \cos^2 \theta$, $b =\sin^2 \theta$, we have 
\begin{align}
    P({\tilde K}_{11}) 
    =
    \frac{ 1 }{ \pi } \frac{ 1 }{ {\tilde K}_{11}^2 + 1 }.
\end{align}
 The same is true for ${\tilde K}_{22}$.

The off-diagonal element of the short-range $K$-matrix is distributed quite differently.  
\begin{align}
    {\tilde K}_{12} 
    =
    \frac{ 1 }{ 2 } \sin(2 \theta) (t_2 - t_1),
\end{align}
with  $\theta$  distributed uniformly through $\theta \in [-\pi/2,\pi/2]$.  The pdf for $u = \sin (2\theta) /2$ ($u \in [-1/2,1/2]$) is given by
\begin{align}
    P(u) 
    &= 
    \frac{ 1 }{ \pi } \times \Big| \cos 2 \theta \Big|^{-1} \nonumber\\
    &= 
    \frac{ 1 }{ \pi } \frac{ 1 }{ \sqrt{ 1 - \sin^2 2\theta} } \nonumber\\
    &= 
    \frac{ 1 }{ \pi }  \frac{ 1 }{ \sqrt{ 1 - 4u^2} }.
\end{align}
Meanwhile, the pdf of the difference $t=t_2-t_1$ is
\begin{align}
    P(t) 
    &= 
    \int_{-\infty}^{-\infty} dt_1 \frac{ 1 }{ \pi } \frac{ 1 }{ t_1^2 + 1 } 
    \frac{1 }{ \pi } \frac{ 1 }{ (t-t_1)^2 +1 } \nonumber\\
    &= 
    \frac{ 2 }{ \pi } \frac{ 1 }{ t^2 + 4 }.
\end{align}
This is another Cauchy distribution, but one with twice the FWHM; this is a special case of \ref{eq:combined_Cauchy}.  Finally, the product is composed to give
\begin{align}
    P({\tilde K}_{12}) 
    &\propto 
    \int_{-1/2}^{1/2} du \frac{ 1 }{ \sqrt{ 1 - 4u^2} }
    \frac{ 1 }{ (u/{\tilde K}_{12})^2 + 4 } \frac{ 1 }{ |u| }, \\
    P({\tilde K}_{12})  
    &= 
    \frac{ 2 }{ \pi^2 } \frac{ 1 }{ \sqrt{ ({\tilde K}_{12})^2 + 1} }
    \sinh^{-1}\left( \frac{ 1 }{ |{\tilde K}_{12}| } \right). 
\end{align}
Here the argument of the inverse hyperbolic sine function makes the distribution divergent at ${\tilde K}_{12}=0$, emphasizing small values of this parameter.  Yet, the divergence is logarithmic, thus maintaining normalizability.

To get to the distributions for the imaginary part of the scattering length requires yet a few more steps.
Given that $x={\tilde K}_{22}$ is Cauchy distributed as above, define $v = 1/(x^2+1)$.  Well, 
\begin{align}
\frac{ dv }{ dx } &= \frac{ 2x }{ x^2 +1 } = 2xv^2, \\
P(v) & \propto v \frac{ 1 }{ xv^2 } \\
& = \frac{ 1 }{ \pi } \frac{ 1 }{ \sqrt{v(1-v)} },
\end{align}
where $v \in [0,1]$.  Similarly, setting $y = {\tilde K}_{12}$ and $w = y^2$, we have
\begin{align}
P(w) = \frac{ 2 }{ \pi^2 } \frac{ 1 }{ \sqrt{ w(w+1)} } \sinh^{-1}\left( \frac{ 1 }{ \sqrt{w} } \right).
\end{align}
In this notation, we have
\begin{align}
\beta = \frac{ 1 }{ {\tilde K}_{22}^2 + 1 } {\tilde K}_{12}^2 = vw.
\end{align} 
Then the distribution of $\beta$ is given formally by
\begin{align}
    P(\beta) 
    &= 
    \int_0^1 dv \frac{ 1 }{ \pi } \frac{ 1 }{ \sqrt{v(1-v) } }
    \frac{ 2 }{ \pi^2 } \frac{ 1 }{ \sqrt{ (\beta/v)(\beta/v+1) } } \nonumber\\
    &\quad\quad\quad\quad \times
    \sinh^{-1} \left( \frac{ 1 }{ \sqrt{\beta/v} } \right) \frac{ 1 }{ v } \\
    &= 
    \frac{ 1 }{ \sqrt{\beta} } \frac{ 2 }{ \pi^3 }
    \int_0^1 dv  \frac{ 1 }{ \sqrt{v(1-v) } }
    \frac{ 1 }{ \sqrt{ v + \beta } } \sinh^{-1} \left( \sqrt{ \frac{ v }{ \beta } } \right). \nonumber
\end{align}
this expression is somewhat intractable, or at least, Mathematica could not seem to tract it.  

We therefore make an approximation.  We regard $\beta$ as fundamentally determined by the factor ${\tilde K}_{12}^2$, as modified somewhat by $v = 1/({\tilde K}_{22}^2+1)$.  The probability distribution for $v$ is seen to be strongly peaked around $v=0$ and $v=1$.  For values of $v$ near unity, ${\tilde K}_{22}^2$ is hardly changed, whereas when $v \approx 0$, the value of ${\tilde K}_{22}^2$ is dramatically reduced.  the influence of $v$ is therefore approximately accounted for by the simplified distribution
\begin{align}
    P^{\prime}(v) 
    =
    \frac{ 1 }{ 2 } \frac{ 1 }{ \sqrt{v} }, \quad v \in [0,1].
\end{align}
With this approximation, the probability distribution for $\beta$ becomes relatively simple:
\begin{align}
    P(\beta) 
    & \approx
        \frac{ 1 }{ \sqrt{\beta} } \frac{ 1 }{ \pi^2 }
    \int_0^1 dv  \frac{ 1 }{ \sqrt{v } }
    \frac{ 1 }{ \sqrt{ v + \beta } } \sinh^{-1} \left( \sqrt{ \frac{ v }{ \beta } } \right). \nonumber \\
    &= \frac{ 1 }{ \pi^2 } \frac{ 1 }{ \sqrt{ \beta } } 
    \left[
    \sinh^{-1} \left( \frac{ 1 }{ \sqrt{ \beta} } \right)
    \right]^2.
\end{align}
This formula does a reasonable job of focusing the probability heavily toward $\beta=0$.


\bibliography{main}

\end{document}